\documentclass[]{tGIS2e}

\usepackage{setspace}
\usepackage{lineno}
\usepackage{multirow}
\usepackage{color,soul}
\usepackage{url}
\usepackage{todonotes}

\usepackage{float}
\floatstyle{plaintop}
\restylefloat{table}

\begin{document}
	
\newcommand\fscore{\mathop{\mbox{$F$-$\mathit{score}$}}}

\doi{10.1080/1365881YYxxxxxxxx}
\issn{1362-3087} \issnp{1365-8816} \jvol{00} \jnum{00} \jyear{2018} 



\title{A natural language processing and geospatial clustering framework for harvesting local place names from geotagged housing advertisements}

\author{Yingjie Hu $^1$, Huina Mao $^2$, and Grant McKenzie $^3$\\
	$^1$ GSDA Lab, Department of Geography, University of Tennessee, Knoxville, TN 37996, USA \\
	$^2$ Geographic Information Science and Technology Group, Oak Ridge National Laboratory, Oak Ridge, TN 37830, USA \\
	$^3$ Department of Geographical Sciences, University of Maryland, College Park, MD 20742, USA \\
	}

\maketitle

\begin{abstract}
\textbf{Abstract:} Local place names are frequently used by residents living in a geographic region. Such place names may not be recorded in existing gazetteers, due to their vernacular nature, relative insignificance to a gazetteer covering a large area (e.g., the entire world), recent establishment (e.g., the name of a newly-opened shopping center), or other reasons. While not always recorded, local place names play important roles in many applications, from supporting public participation in urban planning to locating   victims in disaster response. In this paper, we propose a computational framework for harvesting local place names from geotagged housing advertisements. We make use of those advertisements posted on local-oriented websites, such as Craigslist, where local place names are often mentioned. The proposed framework consists of two stages: natural language processing (NLP)  and geospatial clustering. The NLP stage examines the textual content of housing advertisements, and extracts place name candidates. The geospatial stage focuses on the coordinates associated with the extracted place name candidates, and performs multi-scale geospatial clustering to filter out the non-place names. We evaluate our framework by comparing its performance with those of six baselines. We also compare our result with four existing gazetteers to demonstrate the not-yet-recorded local place names discovered by our framework.

\bigskip

\bigskip

\newpage
\end{abstract}

\pagenumbering{arabic}



\section{Introduction}
Place names play important roles in geographic information science and systems. While computers use numeric coordinates to represent places, people generally refer to places via their names. Digital gazetteers provide organized collections of place names, place types, and their spatial footprints, and fill the critical gap between formal computational representation and informal human discourse \cite[]{hill2000core,goodchild2008introduction,janowicz2008role,kessler2009agenda}. Accordingly, digital gazetteers (hereafter \textit{gazetteers}) are widely used in many applications. 

A number of gazetteers have been developed  by government agencies, commercial companies, and research communities. The Geographic Names Information System (GNIS) is a gazetteer developed by the U.S. Geological Survey and the U.S. Board on Geographic Names, which covers the major place names inside the United States. By contrast, GEOnet Names Server (GNS), developed by the U.S. National Geospatial-Intelligence Agency, is a gazetteer covering place names outside the U.S. Some social media companies, such as Foursquare, have developed their own gazetteers which often focus on points of interest (POI), such as restaurants and stores \cite[]{mckenzie2015poi}. GeoNames is an open  gazetteer which contains over 10 million place names throughout the world (\url{http://www.geonames.org/about.html}). It incorporates gazetteers from multiple countries, such as the U.S. (including GNIS), the U.K., Australia, and Canada, and also contains open data from some commercial companies, such as hotels.com. Who's On First (WOF) (\url{https://whosonfirst.mapzen.com}) is an open gazetteer started by the mapping company Mapzen in 2015, and contains place entries from  Quattroshapes, Natural Earth, GeoPlanet, GeoNames, and the Zetashapes project. WOF selectively merges subsets of place entries from these sources rather than directly combining all of their data \mbox{\cite[]{WOFblog2015}}. The Getty Thesaurus of Geographic Names (TGN) is a gazetteer developed and maintained by the Getty Research Institute, which contains both current and historical place names. There also exist other gazetteers, such as the Alexandria Digital Library Gazetteer (ADL) \cite[]{janee2004issues} and DBpedia Places \cite[]{lehmann2015dbpedia,zhu2016spatial}.   

Some local place names, however, are not recorded in existing gazetteers. There are at least three reasons that can be attributed. First, some place names are vernacular in nature \cite[]{hollenstein2010exploring}. They can be non-standard place names (e.g., ``WeHo" for ``West Hollywood"), abbreviations (e.g., ``BSU" for ``Boise State University"), nicknames (e.g., ``K-Town" for ``Koreatown"), portmanteaus (e.g., ``TriBeCa" for ``Triangle Below Canal Street"), or others. These vernacular places can have vague geographic boundaries that are hard to delineate accurately \cite[]{twaroch2009acquisition}. Thus, while frequently used, vernacular place names are often not officially recorded. Second, some gazetteers are designed to cover a large geographic extent rather than a local area. For example, GNIS aims to cover place names in the entire U.S., and some local geographic features or locally-used names may be considered as relatively ``insignificant" and are thus omitted. Third, keeping a gazetteer up-to-date takes a considerable amount of time and human resources. Consequently, the names of some newly-constructed entities may not be included. 


Local place names have great values to a variety of applications. In disaster response, local place names are often observed in incident reports in short text messages or tweets (whose length limitation also prompts the use of local place names that are often shorter than official names)  \cite[]{gelernter2011geo}. Meanwhile, disaster response teams can come from other cities, states, or even countries, and may not be familiar with the place names used by local residents. A gazetteer containing local place names, thus, can help automatically interpret the incident reports and locate the people in need. Local place names can also be used in public participation GIS (PPGIS) \cite[]{rinner2009evaluating,hu2015multistage,kar2016public}, especially its application in urban planning. Consider a scenario in which both professionals and local residents are engaged in a public meeting to discuss a city planning project. Residents may use local place names to refer to certain local areas. A PPGIS, with the capability of understanding and locating these local place names, can facilitate the discussion between professionals and residents \mbox{\cite[]{brown2015engaging}}. Local place names can be useful in other applications as well, such as locating transitory obstacles by geoparsing volunteer-contributed text messages to assist blind or vision-impaired pedestrians \cite[]{rice2012supporting,aburizaiza2016geospatial}.  


This paper proposes a computational framework for harvesting local place names which can be used for enriching gazetteers. Specifically, we make use of geotagged housing advertisements posted on local-oriented websites, such as Craigslist (\url{https://www.craigslist.org}). Our main contributions are twofold:
\begin{itemize}

\item From a methodological perspective, this paper contributes a two-stage computational framework that integrates natural language processing and geospatial clustering for harvesting local place names. 

\item From an application perspective, this paper proposes an innovative use of geotagged and local-oriented housing advertisements on the Web for extracting local place names and enriching gazetteers.


\end{itemize}

The remainder of this paper is organized as follows. Section 2 reviews related work on place name extraction, disambiguation, and gazetteer enrichment. Section 3 presents our framework, and explains the methodological details of the two-stage process. Section 4 applies the proposed framework to an experimental dataset of geotagged housing advertisements collected from six different geographic regions, and discusses the experiment results. Finally, section 5 summarizes this work and discusses future directions.





\vspace*{-0.3cm}
\section{Related work}
Place names (or \textit{toponyms}) are widely used in various types of texts, such as news articles \cite[]{lieberman2011multifaceted,liu2014analyzing}, travel blogs \cite[]{leidner2011detecting,adams2012geo}, social media posts \cite[]{kessler2009bottom,zhang2014geocoding}, housing advertisements \cite[]{medway2014s,madden2017pushed}, historical archives \cite[]{southall2014rebuilding,delozier2016creating}, Wikipedia pages \cite[]{hecht2008geosr,salvini2016spatialization}, and others \cite[]{gregory2015geoparsing}. Recognizing place names from texts and linking them to spatial footprints are important steps for automatically understanding the semantics of natural language texts, and are studied in both computer science and GIScience \cite[]{larson1996geographic,mccurley2001geospatial,jones2008geographical,vasardani2013locating,karimzadeh2013geotxt,melo2017automated,doi:10.1080/13658816.2017.1368523}. 

Gazetteers, as geographic knowledge bases, are frequently used for the task of place name recognition. One straightforward usage is to determine the qualification of a word or a phrase as a place name, which is often done by checking its existence in a gazetteer \cite[]{li2002location,stokes2008empirical,lieberman2011multifaceted}. A more advanced usage of gazetteers is place name disambiguation (or \textit{toponym resolution}). Since multiple place names can refer to the same place instance and the same place name can refer to different place instances, it is challenging to determine which place instance was referred to by a name in the text \cite[]{amitay2004web,leidner2008toponym,hu2014improving}. Gazetteers have been used in many ways for supporting place name disambiguation. Based on the related places in a gazetteer (e.g., higher administrative units), researchers  developed methods, such as co-occurrence models \cite[]{overell2008using} and conceptual density \cite[]{buscaldi2008conceptual}, to disambiguate the mentioned place names. Based on the spatial footprints of place instances, researchers  designed heuristics  for place name disambiguation, e.g., place names mentioned in the same document generally share the same geographic context  \cite[]{leidner2008toponym,lieberman2010geotagging,paradesi2011geotagging,santos2015using,awamura2015location}. The metadata of places contained in a gazetteer, such as population, are also used for disambiguation, e.g., by assigning prominent instances as the default senses of place names or using metadata as additional features to determine the correct place instances \cite[]{li2002location,ladra2008toponym,zhang2014geocoding}. Some place name recognition methods were designed without using a gazetteer. For example,  \cite{adams2012geo} and \cite{delozier2015gazetteer} statistically summarized the geographic distributions of words over the surface of the Earth using Wikipedia and travel blog articles. Such geographic distributions can be utilized for disambiguating a target place name based on its context words. \cite{inkpen2015location} used both a gazetteer and word features (e.g., part of speech, left words, and right words) to train a conditional random field model which can extract cities, states, and countries from texts.

Many other studies focused on enriching gazetteers with additional information. One important topic is representing the vague boundaries of vernacular places so that they can be added to a gazetteer. \cite{montello2003s} identified the common core area of ``downtown Santa Barbara" by inviting human participants to draw the boundaries of downtown in their beliefs on a map. \cite{jones2008modelling} used a Web search engine to harvest geographic entities (e.g., hotels) related to a vague place name (e.g., ``Mid-Wales"), and utilized the locations of these harvested entities to construct the vague boundary. Flickr photo data present a natural link between textual tags and locations, and are used in many studies on identifying boundaries for vague places \cite[]{grothe2009automated,kessler2009bottom,intagorn2011learning,li2012constructing}.  

Existing studies, however, often assume that a place name is already given and the task is to construct the best spatial footprint for this place name. In this work, we examine a different question, namely \textit{given a geographic region, what are the local place names used by residents there but not yet recorded in gazetteers?} Some researchers have looked into this problem. \cite{twaroch2010web} developed a Web-based platform, called ``People's Place Names" (\url{http://www.yourplacenames.com}), which explicitly invites local people to contribute vernacular place names. While such a platform is useful, it can be challenging to constantly encourage people to contribute, especially over a long time period. In another study, \cite{gelernter2013automatic} proposed a matching algorithm which can compare the tags in OpenStreetMap and Wikimapia with the place entries in a gazetteer, and can add the place information that are not contained in a gazetteer. Our work aligns with the general direction of these two studies, but utilizes geotagged housing advertisements posted on local-oriented websites for harvesting local place names. In the following, we present our methods and describe the advantages of using geotagged housing advertisements for collecting local place names.



\section{Methods}
\subsection{Overall architecture}
\begin{figure}[h]
	\begin{center}
		\includegraphics[width=\textwidth]{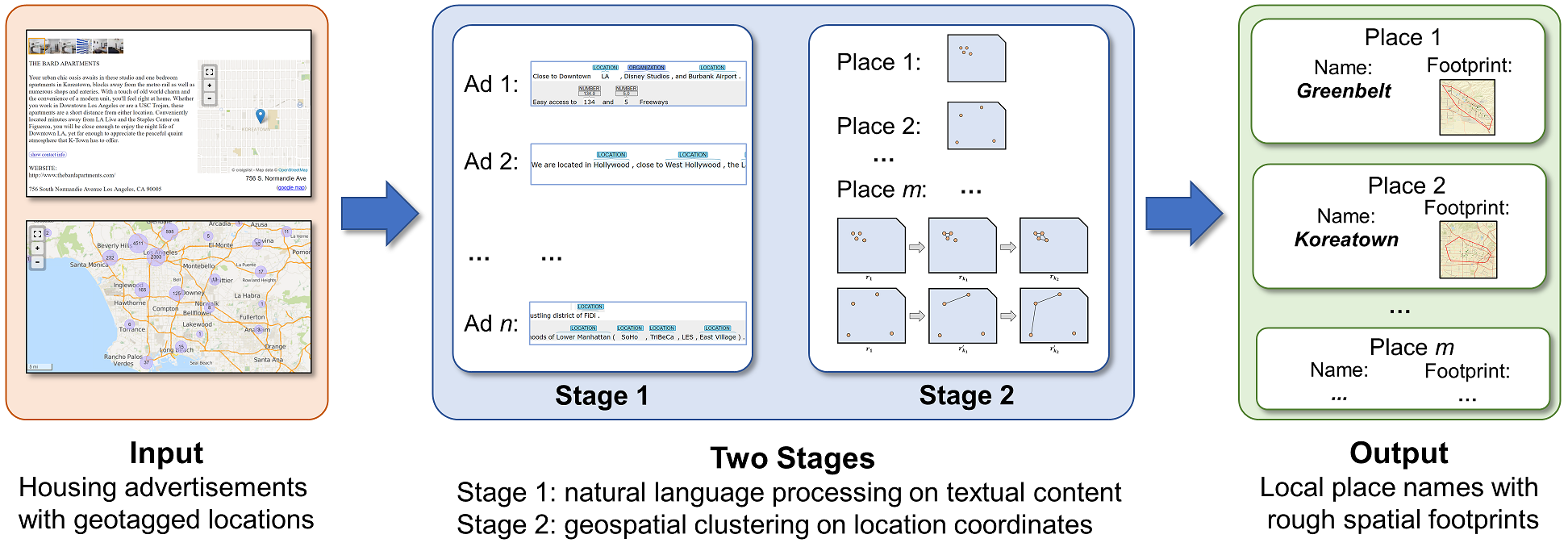}
		\caption{Overall architecture of the proposed two-stage framework.}
		\label{framework}
	\end{center}
\end{figure}

We develop a two-stage computational framework which takes the geotagged housing advertisements from a target geographic region as the input, and outputs the identified local place names and their rough spatial footprints. Figure \ref{framework} shows the overall architecture of this framework. 


\vspace*{-1.3cm}
\subsection{Input: geotagged housing advertisements} 
One unique feature of the proposed framework is the use of geotagged housing advertisements posted on local-oriented websites. In this work, a geotagged housing advertisement is an advertisement tagged with the location (a latitude-longitude pair) of the advertised housing property. This type of data is available in many housing websites nowadays. For housing advertisements without geotagged locations, it is possible to assign coordinates to them by geocoding the addresses of the advertised properties. There are several advantages in using housing advertisements for extracting local place names. First, local place names are often mentioned in these advertisements. Location is commonly recognized as the most important factor in making housing decisions. Thus, writers of housing advertisements are fully motivated to demonstrate the location convenience of the advertised property by describing its neighborhood and nearby facilities, and local place names are often used in these descriptions. Second, housing advertisements can be found in many geographic areas where people live, and often have digital versions online. This increases the applicability of the proposed framework: to harvest local place names in an area, we can first collect the housing advertisements in that area (e.g., by crawling local housing websites), and then apply our framework to the collected data. Finally, housing advertisements can help discover newly-established place names, since they are posted constantly. 

Local place names also exist in other data sources, such as social media. However, such data often contain too much noise and cannot be directly used for collecting local place names. For example, a tweet geotagged to a neighborhood can be talking about any topics, not necessarily related to the local neighborhood. In addition, a user can mention a place from almost anywhere without having to physically stay there. 
While data from Flickr, a photo sharing website, present a stronger connection between texts and locations than tweets, they often reflect the perspectives of tourists rather than of local people \cite[]{girardin2008digital}. Data from Instagram also contain a lot of noise. Due to these limitations of social media data, we use geotagged housing advertisements  as the input for the proposed framework.


\vspace*{-0.5cm}
\subsection{Stage 1: Natural language processing}
Each geotagged housing advertisement in the input dataset consists of two parts: a textual description and a geographic location. The first stage of our framework examines the textual descriptions of the advertisements. The goal is to identify as many place names as possible from these descriptions. From a perspective of information retrieval, this stage aims to increase the \textit{recall} of the extracted place names.

A major challenge of Stage 1 is that we cannot use an existing gazetteer (or any methods that purely rely on gazetteers) to extract place names. This is because the goal of this work is to identify the local place names that are not yet recorded in gazetteers. Accordingly, we resort to natural language processing (NLP) models which can extract place names beyond those in a gazetteer. Since false positives (non-place names) can also be included by NLP models, we consider their output as \textit{place name candidates}. Another challenge lies in the informal format of housing advertisements, especially those posted by individuals on local websites. For example, some housing advertisements use capital letters for the entire posts (e.g., ``BEAUTIFUL STUDIO IN DOWNTOWN BOISE ..."), while some use capital letters to emphasize certain phrases (e.g.,``This apartment has a HUGE bedroom."). In these situations, the performance of a NLP model trained using well-formated texts (e.g., news articles) can be limited.



\begin{figure}[h]
	\begin{center}
		\includegraphics[width=\textwidth]{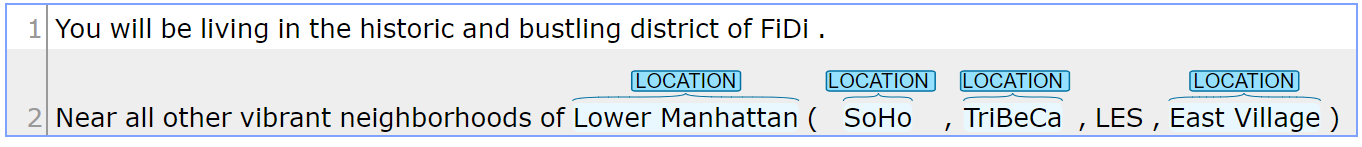}
		
		\caption{An example of named entity recognition  using the default Stanford NER model.}
		\label{NER_example}
	\end{center}
	\vspace*{-0.5cm}
\end{figure}

To address the two challenges, we use a combination of off-the-shelf and retrained named entity recognition (NER) models. 
The input of a NER model is the textual description of a housing advertisement, and the output is the text with annotated entities. Figure \ref{NER_example} shows an example of identifying locations from two sentences of a housing advertisement in New York City using the default (off-the-shelf) Stanford NER model. 
As can be seen, place names, such as ``Lower Manhattan", ``SoHo", and ``TriBeCa", are identified, while two other place names, ``FiDi" (Financial District) and ``LES" (Lower East Side),  are missed by this default model. To identify as many place name candidates as possible, we make use of four NER models: spaCy NER, default Stanford NER, case-insensitive Stanford NER, and Twitter-retrained Stanford NER. In the following, we provide more details about each of them. 


\noindent \textbf{1) spaCy NER.}
spaCy (\url{https://spacy.io/}) is an open source software library for natural language processing  in Python and Cython. spaCy NER uses linear models for named entity recognition, with weights learned using the averaged perceptron algorithm. It identifies PERSON, NORP (e.g., nationalities and political groups), FACILITY (e.g., buildings, airports, and highways), ORG (e.g., companies, agencies, and institutions), GPE (e.g., countries, cities, and states), LOC (e.g., non-GPE locations, mountain ranges, and bodies of water), and other types of entities. 
 spaCy NER is trained on the OntoNotes 5.0 corpus (\url{https://catalog.ldc.upenn.edu/LDC2013T19}) using the part-of-speech (POS) tag and Brown cluster of words as training features. Given our interest in place names,  we keep only FACILITY, ORG, GPE, and LOC in the extracted entities.
  

\noindent \textbf{2) Default Stanford NER.} Compared to spaCy NER which started in 2014, Stanford NER has been used for over a decade, with its first release in 2006 followed by multiple updated versions (\url{https://nlp.stanford.edu/software/CRF-NER.shtml}). Stanford NER is one of the state-of-the-art tools, which uses conditional random field (CRF) models and distributional similarity features to improve entity recognition accuracy and efficiency \citep{finkel2005incorporating}. The training features of Stanford NER include word features (e.g., current and surrounding words), orthographic features, prefixes and suffixes, POS tags, and lots of feature conjunctions. A CRF is a sequence model that aims to find the most likely state sequence given some observations~\cite[]{lafferty2001conditional}. In the task of NER, observations are a sequence of words, and the states to be found are a sequence of entity tags. Let $\mathbf{x} = \{x_0, x_1, ..., x_n\}$ represent a sentence ($x_i$ represents a word), and let $\mathbf{y} = \{y_0, y_1, ..., y_n\}$ represent the corresponding entity tags of the words. The probability of $\mathbf{y}$ given $\mathbf{x}$ can be calculated using Equation \ref{ner_stanford}:
\begin{equation} \label{ner_stanford}
P_{\mathit{crf}} (\mathbf{y}|\mathbf{x}) \propto   \prod_{i=1}^{n} \phi (y_{i-1}, y_i )
\end{equation}
where $ \phi (y_{i-1}, y_i )$ is the probability between an adjacent pair of states at positions $i-1$ and $i$. Based on this equation, the Viterbi algorithm \cite[]{forney1973viterbi} is used to infer the most likely state sequence.  A major advantage of using CRF for detecting named entities is that each word is not treated independently but is considered within a sequence. Stanford NER has three-class (i.e.,~LOCATION, PERSON, ORGANIZATION), four-class, and seven-class models. 
In this work, we use the three-class model and keep only LOCATION and ORGANIZATION in the extracted result.  

\noindent \textbf{3) Case-insensitive Stanford NER.} The default Stanford NER model was trained using well-formatted text data, such as CoNLL 2003 \cite[]{tjong2003introduction}. 
As discussed previously, housing advertisements posted on local websites are often written in informal formats. To better detect local place names, we employ the case-insensitive version of Stanford NER which ignores the case of words and was trained using only lowercase texts. 


 
\noindent \textbf{4) Twitter-retrained Stanford NER.}
Case-insensitive Stanford NER can help identify place names from the descriptions that are informally capitalized. However, it was still trained based on relatively well-structured sentences with subject, predicate, and object, and with mostly formal word spelling. In a local housing advertisement, one sentence can be followed by more than one exclamation marks (e.g., ``An Apartment You Must See!!!"), may contain abbreviations and irregular spellings (e.g., ``asap" and ``The price is soooooo low!"), or may omit part of the subject-predicate-object structure (e.g., ``Great location in NoHo."). Previous research has shown that retraining NER models using annotated informal texts can significantly boost their performances in similar text environments \cite[]{lingad2013location}. In this work, we retrain the default Stanford NER model using a human annotated Twitter dataset from the ALTA 2014 Twitter Location Detection shared task \citep{molla2014overview}. 



With the four NER models prepared, we take a union strategy by applying them to the same housing advertisement and combining the extracted place name candidates. In the Experiments section later, we will systematically evaluate the performances of the four individual models, as well as the performances of the combined models.

\subsection{Stage 2: Geospatial clustering}
Stage 1 identifies place name candidates which also contain false positives. A major reason for this result is because the  NER models have to tolerate many variations and irregularities of the local place names mentioned in housing advertisements, such as ``Nolita" and ``K-Town". Besides, place names do not necessarily follow prepositions like ``in" or ``at", especially given the informal language in local housing advertisements. To accommodate these various situations, the NER models inevitably include words and phrases that are not place names. The goal of Stage 2, therefore, is to filter out as many of these false positives as possible. From a perspective of information retrieval, Stage 2 aims to increase the \textit{precision} of the extracted place names.   

The main data examined in Stage 2 is the location coordinates associated with the place name candidates. In the output of Stage 1, each place name candidate is linked to a number of points which are the geotagged locations of the housing advertisements that mention this particular place name candidate. In Stage 2, we analyze the distribution patterns of these coordinates to identify the true place names. Intuitively, the coordinates associated with a true place name, such as ``K-Town", are more likely to show a geospatial cluster, since it is often mentioned in advertisements whose housing properties are located in or near these areas. In contrast, a non-place name, such as ``Central AC" (the linguistic pattern of this phrase is, in fact, similar to a true place name, such as ``Downtown LA"), can show up in almost any housing advertisements, and the associated locations are more likely to be scattered around the entire study region. Based on this intuition, we formalize the task of Stage 2 as a geospatial clustering problem. However, one critical challenge is that the clusters can be at different geographic scales.  For example, the coordinates associated with ``K-Town" may form a cluster at the neighborhood scale, while the coordinates associated with ``Towne Square Mall" may form a cluster at a point-of-interest scale. Examining the coordinates of ``K-Town" at the point-of-interest scale may not reveal a cluster. Thus, we cannot use the clustering methods which detect clusters based on a single distance value.

\begin{figure}[h]
	\vspace*{-0.5cm}
	\begin{center}
		\includegraphics[width=0.9\textwidth]{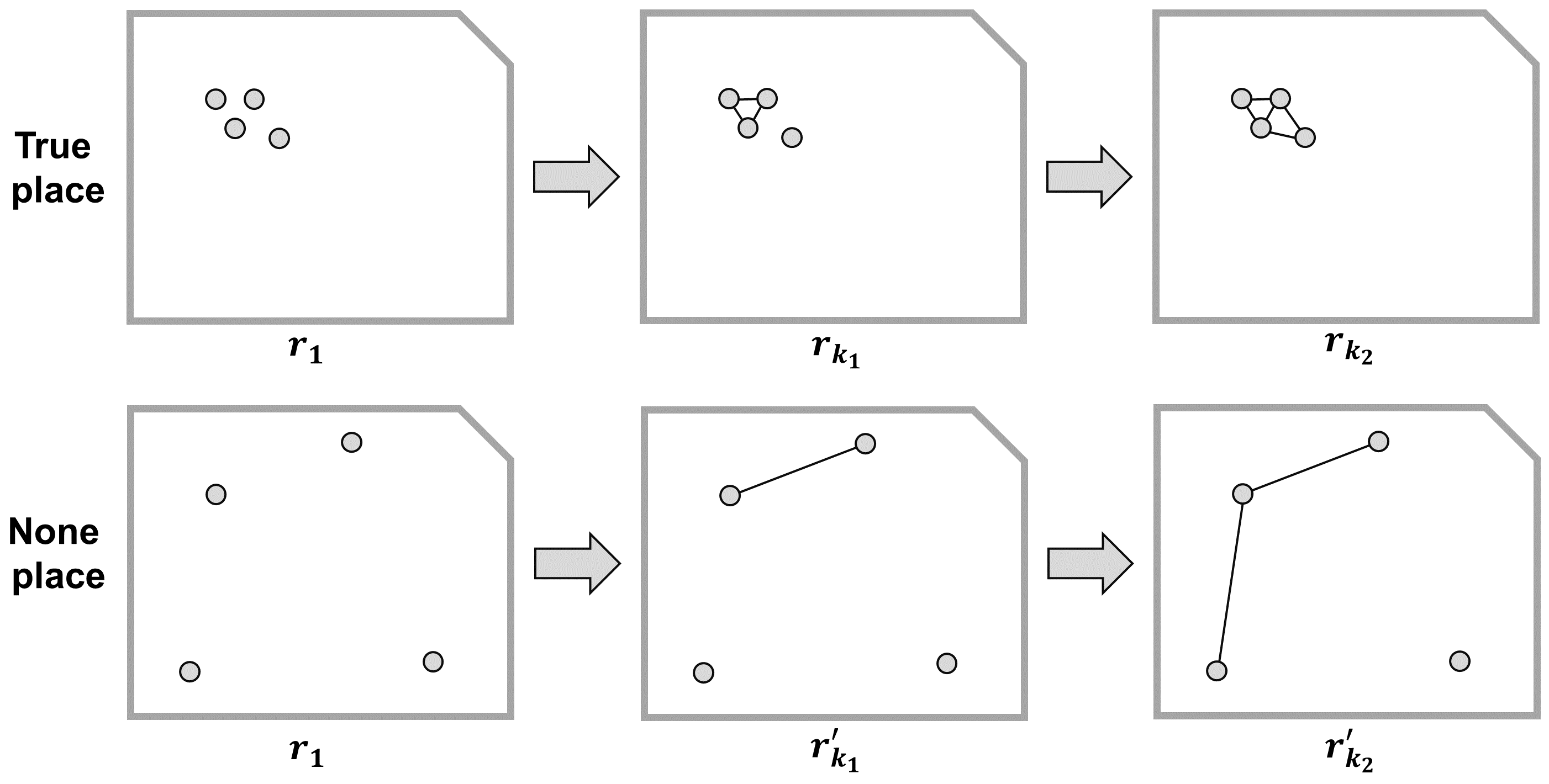}

		\caption{Illustration of the scale-structure identification algorithm.}
		\label{ssi_illus}
	\end{center}
\vspace*{-0.5cm}
\end{figure}

To address this challenge, we employ and modify the scale-structure identification (SSI) algorithm to rank the \textit{geo-indicativeness} of the place name candidates. SSI algorithm was initially proposed by \cite{rattenbury2007towards} from Yahoo! Research to identify the place semantics of Flickr tags. 
It attempts to cluster point coordinates at multiple geographic scales and examines their overall ``clusterness", and therefore can overcome the challenge that coordinates may form clusters at different scales. In the following, we briefly explain the mechanism of SSI. Let $x$ represent a place name candidate (a \textit{term} for short), and let $L_x$ represent a set of points associated with $x$. SSI functions as follows: 1) let $R = \{r_k | k=1,2,3, ...,K\}$ be an ordered set of distances that define the multiple clustering scales, and $r_k = \alpha^k, \alpha > 1$ (we use $\alpha = 2$ meters 
in this work); 2) consider the points in $L_x$ as the nodes of a graph, calculate the pair-wise distances between all points, and let $d_{ij}$ represent the distance between point $i$ and $j$; 3) iterate $k$ from $1$ to $K$, and at each distance threshold $r_k$, build an edge between point $i$ and $j$ if $d_{ij} <= r_k$; 4) calculate the entropy $E_k$ of the graph at scale $k$ using Equation \ref{SSI_equation1}:
\begin{equation} \label{SSI_equation1}
E_k = -\sum_{y \in Y_k}\frac{|y|}{|L_x|} log\frac{|y|}{|L_x|}
\end{equation}    
where $Y_k$ represents a set of connected components of the graph under scale $k$, and $y$ represents a connected component. $|y|$ is the number of points in this connected component, and $|L_x|$ represents the total number of points associated with term $x$; 5) finally, the geo-indicativeness of the term $x$ is quantified by summing up $E_k$ at all scales: $\sum_{k} E_k$. Figure \ref{ssi_illus} illustrates SSI by comparing the clustering processes of a true place name and a non-place one.

In Fig. \ref{ssi_illus}, the gray outline represents the target geographic area (e.g., the Greater Los Angeles Area). As can be seen, the points associated with a true place name, e.g., ``K-Town", tend to cluster at a sub region of the study area, while the points associated with a non-place name, e.g., ``Central AC", can be scattered around the entire area. 
When SSI starts from $r_1$ (e.g., a distance of $2$ meters), all nodes of both the true place and none place examples are disconnected, and thus each single node is an individual component. As $r_k$ increases, the nodes of a true place quickly become connected and eventually form one single connected component. By contrast, the nodes of a none place only  connect slowly as $r_k$ increases. Note that $r_1 < r_{k_1} < r_{k_2} < r_{k_1}' < r_{k_2}'$. If we calculate the sum of entropies at all the scales, the true place will have a smaller entropy sum than that of the none place, since after the scale of $r_{k_2}$ all entropies become $0$. 
Thus, we can rank place name candidates based on their entropy sums in an ascending order, and true place names should show up at higher ranks. 



\begin{figure}[h]
	\begin{center}
		\includegraphics[width=0.65\textwidth]{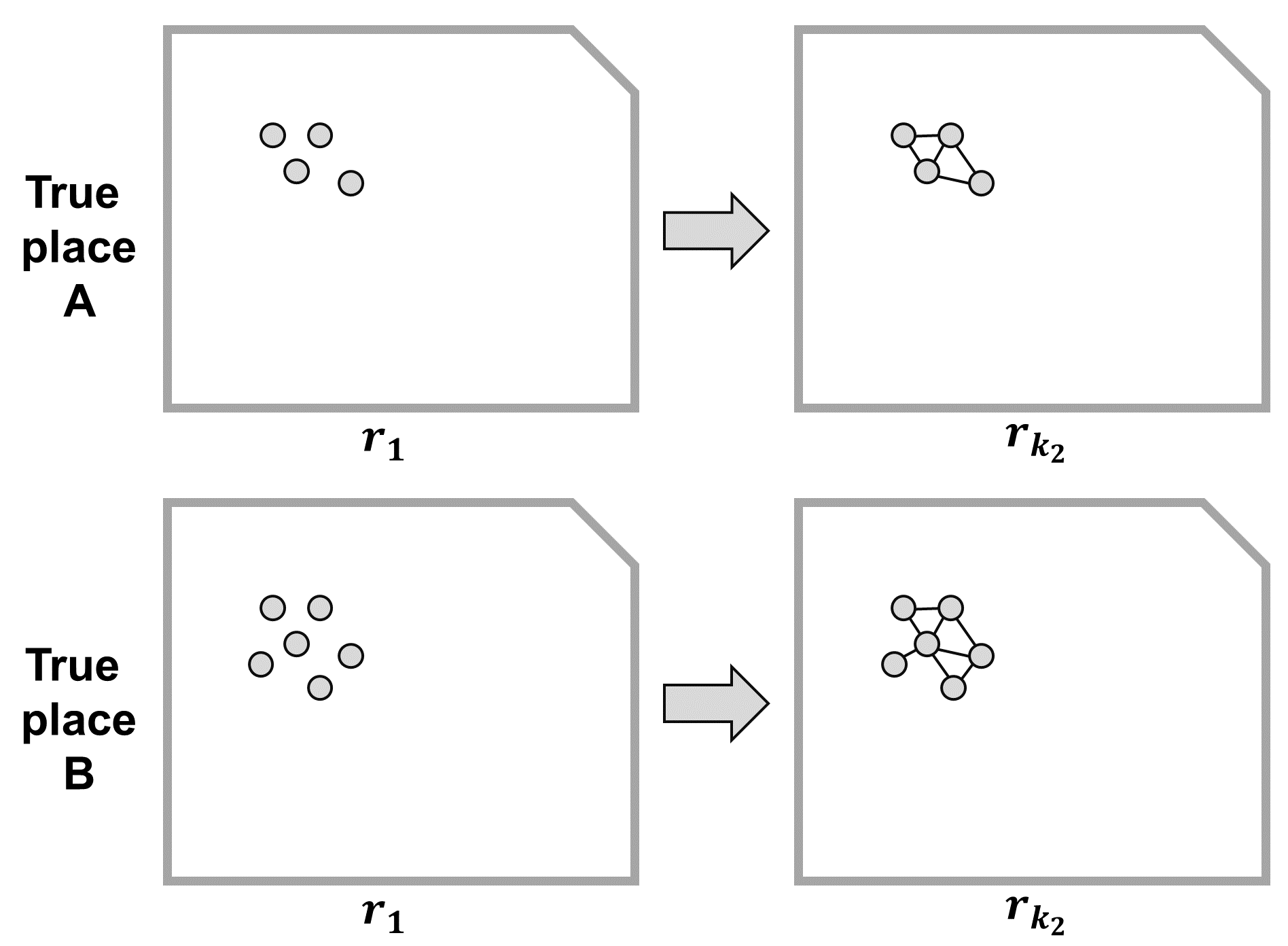}
		\caption{Illustration of a limitation of the scale-structure identification algorithm.}
		\label{ssi_illus2}
	\end{center}
\vspace*{-2cm}
\end{figure}

While SSI is theoretically sound, our pilot experiments identified a limitation of this algorithm when the number of points associated with a term is small (e.g., fewer than $10$). Consider the example in Fig. \ref{ssi_illus2} in which both $A$ and $B$ are true place names, but B has more points than $A$. 
It can be seen that the points of $A$ and $B$ are distributed in a similar geographic pattern, and both become one single connected component under the same distance threshold $r_{k_2}$. In an ideal case, $A$ and $B$ should have similar geo-indicativeness. However, the current SSI penalizes the place names associated with more points. In Fig. \ref{ssi_illus2}, the entropy of true place $A$ at $r_1$ is $-\sum_{i=1}^{4} (\frac{1}{4} \times \log\frac{1}{4}) = 2.00$, whereas the entropy of true place $B$ at $r_1$ is $-\sum_{i=1}^{6} (\frac{1}{6} \times \log\frac{1}{6}) = 2.59$. Thus, $B$ has a higher entropy than $A$ based on existing SSI. 
This is not a problem in the original work by \cite{rattenbury2007towards}, since in their dataset one Flickr tag is associated with in average $232.26$ points (e.g., $\log\frac{1}{232}=-7.858, \log\frac{1}{233}=-7.864$). In our case, a lot of place name candidates are associated with fewer than 10 points. To mitigate this issue, we modify the existing SSI into Equation \ref{extended_SSI}.
\begin{equation} \label{extended_SSI}
E_x =  \frac{1}{\sqrt{|L_x|}}\sum_{k=1}^{K} E_k
\end{equation} 
where 
the original sum of entropies $\sum_{k=1}^{K} E_k$ is adjusted based on the number of points. The square root dampens the effect of the point count, and helps ensure that point count does not dominate the entropy sum. This square-root adjustment is determined empirically, and we also tested other approaches, such as $1/|L_x|$ and $1/\log{|L_x|}$. We will present the empirical comparisons in the Experiments section. With our modified SSI, place $A$ and place $B$ now have entropy sums, $1.00$ and $1.06$, which are more similar. This modified SSI can also be considered as modeling two factors: the degree of clusterness and the count of endorsements. The terms, which are highly clustered and which are endorsed by many advertisement writers (each mention can be seen as one endorsement), are more likely to be true place names. 


\vspace*{-0.5cm}
\subsection{Output: Place names with rough spatial footprints}
\begin{figure}[]
	\begin{center}
		\includegraphics[width=\textwidth]{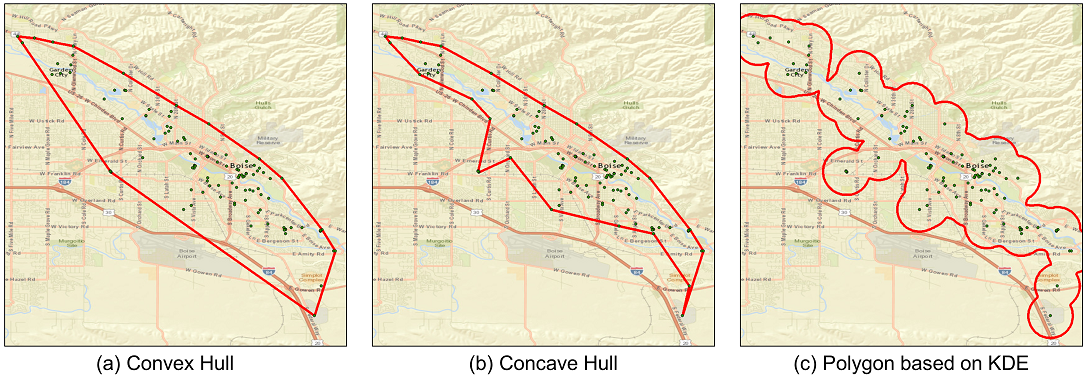}
		\caption{Three methods for constructing rough spatial footprints for the place name, ``Greenbelt", based on the associated housing advertisement locations.}
		\label{spatialfootprint}
	\end{center}
\end{figure}

Stage 2 ranks the geo-indicativeness of the place name candidates based on their entropy sums calculated using the modified SSI. We can then define a threshold, and return place names whose entropy sums are lower than this threshold. Such a threshold can be determined based on precision-recall curves, which will be demonstrated in the following section. In addition, since each place name is associated with a number of point locations, we can construct rough spatial footprints for the extracted place names. A number of methods, such as convex hull \cite[]{jarvis1973identification}, concave hull \cite[]{duckham2008efficient}, and kernel density estimation (KDE) \cite[]{sheather1991reliable}, have been used in previous research for constructing spatial footprints \cite[]{jones2008modelling,li2012constructing,mckenziejuxtaposing}. Figure \ref{spatialfootprint} shows three polygons created based on the point locations associated with the term ``Greenbelt" in Boise, Idaho, USA, using three different methods. We can then choose a method that fits the need of a project. Here, we only demonstrate the feasibility of constructing rough spatial footprints for the extracted place names. Identifying the suitable parameters for delineating the best spatial footprint for a place name is beyond the scope of this work and is worth further investigation.



\vspace*{-0.5cm}
\section{Experiments}
In this section, we apply the proposed two-stage framework to a dataset of geotagged housing advertisements in six different geographic regions to extract local place names. We first describe the dataset and then present the multiple experiments conducted for evaluating the performance of the framework.

\subsection{Dataset}
The experimental dataset was collected from Craigslist which is a local-oriented advertisement website. There is one Craigslist website instance for each geographic region defined by Craigslist. We selected $6$ different regions that contain $6$ U.S. cities, which are: New York City (NY), Los Angeles (CA), Chicago (IL), Richmond (VI), Boise (ID), and Spokane (WA). These regions were selected based on the population rankings of the contained major cities: the former three cities rank as top 3 among all U.S. cities, while the latter three rank as 98th, 99th, and 101th (the city that ranks as 100th is San Bernardino, which is a California city close to Los Angeles; thus we replaced it with Spokane). These 6 regions are in 6 different U.S. states, and the housing advertisements are retrieved from $6$ Craigslist websites respectively.

\begin{figure}[h]
	\begin{center}
		\includegraphics[width=\textwidth]{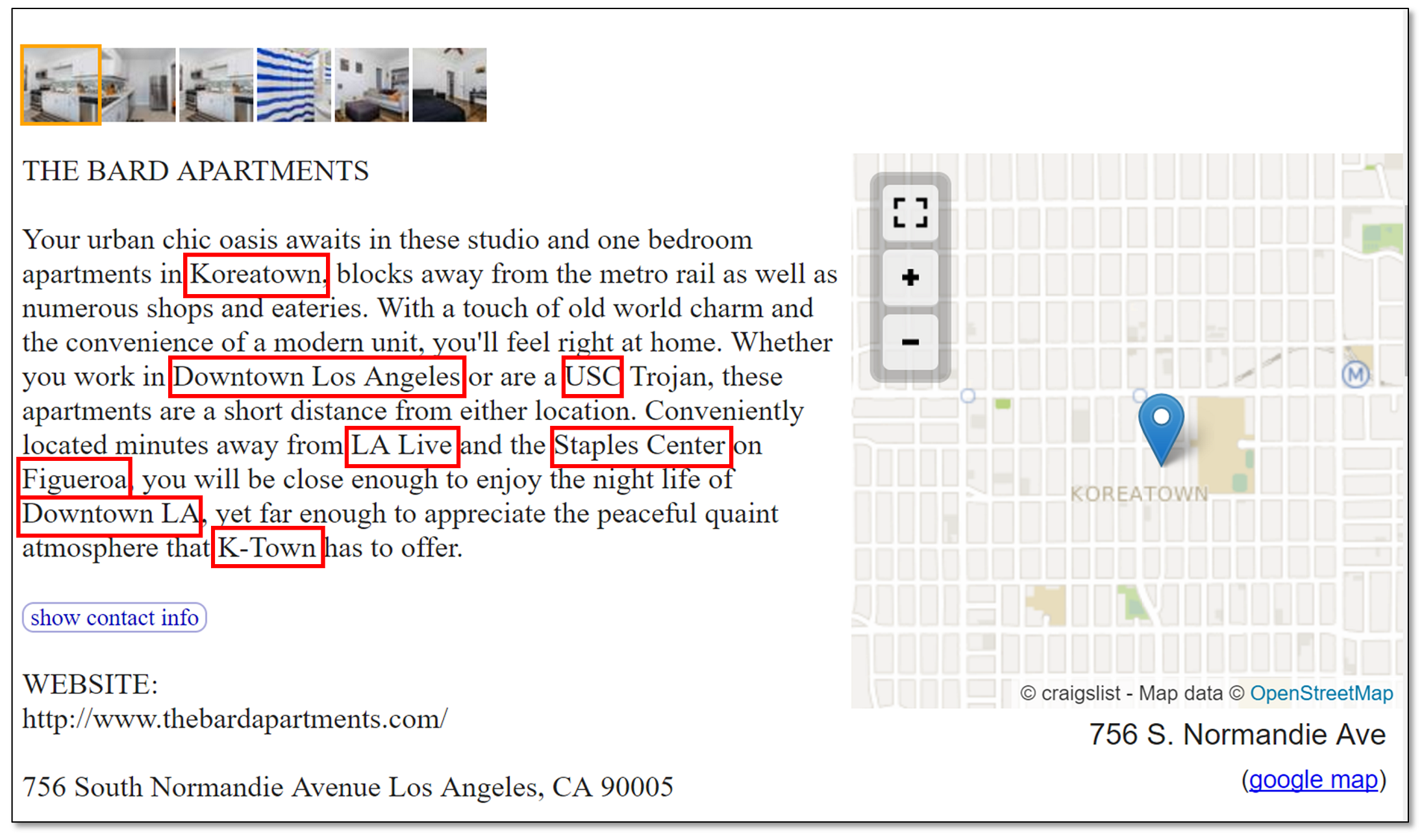}
		
		\caption{A geotagged housing advertisement from a Craigslist website.}
		\label{craigslistPost}
	\end{center}
\end{figure}

The data collection took about three and a half months (from Feb. 18th, 2017 to May 30th, 2017). A Java Web crawler was developed using the library of HtmlUnit to retrieve housing advertisements from Craigslist websites. Figure \ref{craigslistPost} shows an example of a geotagged housing advertisement on the Los Angeles website of Craigslist. As can be seen, the left side of the advertisement provides a textual description on the housing property, which mentions multiple place names including local place names such as ``K-Town". On the right side, a map shows the location of the housing property. Our Web crawler extracts the textual description and the latitude and longitude of the housing location embedded in the HTML page, and no additional geocoding is involved. A retrieved housing advertisement contains the post ID, repost ID (if this is a repost), post time, longitude, latitude, and post content. Some advertisements do not provide location coordinates and are not used in our experiments. In total, we collected more than $2$ GB data with over 3 million housing advertisements for the six study regions. The collected data are stored in individual comma-separated values files.

\begin{table}[h]
	\centering
	\caption{Counts of the distinct geotagged housing advertisements in the 6 study regions.}
	\label{table_adcount}
	\vspace*{0.3cm}
	\begin{tabular}{|c|c|c|c|c|c|}
		\hline
		New York & Los Angeles & Chicago & Richmond & Boise & Spokane \\ \hline
		6205     & 9301        & 8973    & 4712     & 3373  & 3288    \\ \hline
	\end{tabular}
\end{table}

\begin{figure}[H]
	\begin{center}
		\includegraphics[width=\textwidth]{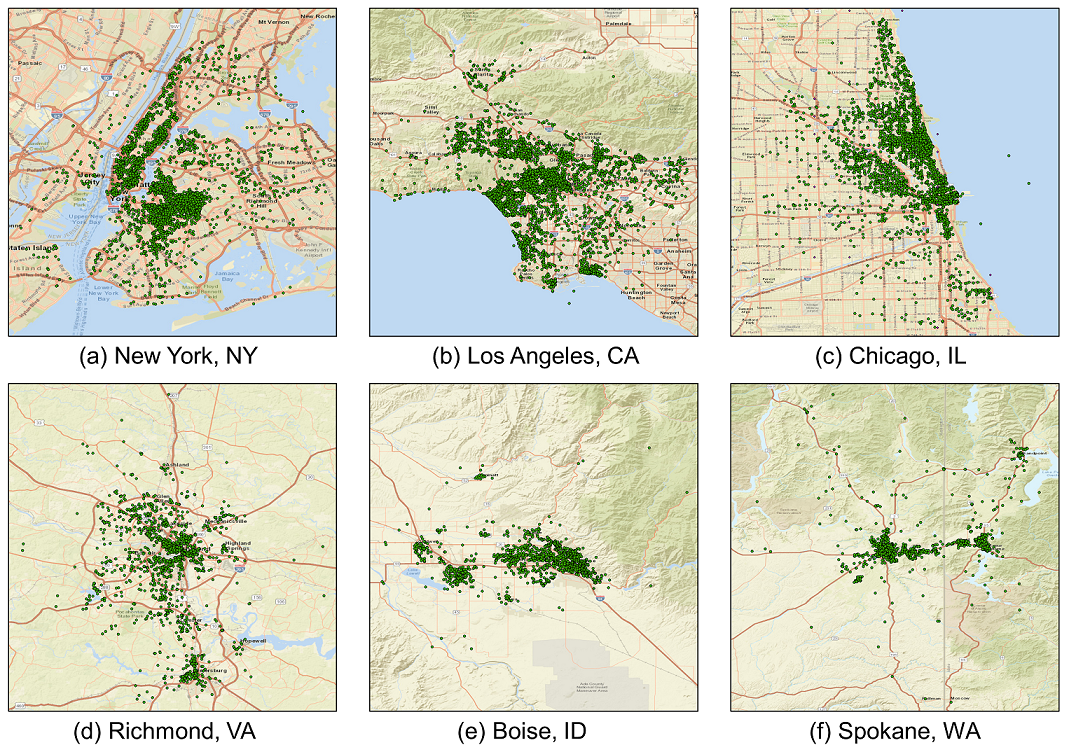}
		\caption{Geographic distributions of the distinct housing advertisements in the 6 study regions.}
		\label{sixcities}
	\end{center}
\end{figure}

The raw data contain a lot of duplications, since a user tends to repost the same advertisement if no response is received after a few days. In addition, some apartment rental companies post a large number of advertisements which dominate the raw data. To reduce the potential bias, we remove the advertisements which are reposts (based on their repost ID) and those whose first 50 characters exactly overlap with the existing posts. These two filters removed a large number of posts, and the final counts of distinct advertisements are summarized in Table \ref{table_adcount}. Figure \ref{sixcities} shows the geographic distributions of the distinct housing advertisements.

\vspace*{-0.5cm}
\subsection{Experiment Procedure} 
With the distinct geotagged housing advertisements, we first apply the four NER models in Stage 1 to the textual description of each post. We combine the place name candidates extracted, and associate each name with the locations of the housing advertisements that mention this name. After Stage 1, we obtain a set of place name candidates for each of the six study regions, and each name is associated with a number of point locations.

In Stage 2, we examine the geo-indicativeness of the place name candidates based on the associated point locations. Before running the modified SSI algorithm, we first apply a spatial filter to the locations to reduce the noise contained in Craigslist data. This is because some advertisements may tag the housing properties with wrong locations, and sometimes an advertisement writer may overly exaggerate the location convenience of the housing property (e.g., by saying that the property is close to a shopping mall even though it is in fact far away). We perform the following operations to reduce the data noise: 1) identify the medoid of all the points associated with a place name candidate (the medoid is identified by first calculating the Euclidean distance between every point pair and then selecting the point with the smallest sum of distances to all other points); 2) find the third quartile distance based on the distances from all other points to the medoid; 3) remove the points whose distances to the medoid are larger than the third quartile distance. These three steps preserve the majority of points close to the medoid, and reduce the number of noise points included. We also remove the place name candidates associated with fewer than 3 points after this filtering process. The modified SSI algorithm is then applied to the data, and the place name candidates are ranked based on their adjusted entropy sums in an ascending order.  

After the two stages, we have obtained a ranked list of place name candidates for each region. We can then determine a threshold for the adjusted entropy sum to identify the candidates that will be considered as true place names. The process of determining the entropy threshold will be discussed in the following subsection. With the identified place names, we can construct rough spatial footprints for them using methods such as convex hull.

\vspace*{-0.5cm}
\subsection{Performance evaluation}
In this subsection, we evaluate the performance of the proposed two-stage framework. We start the evaluation by obtaining a ground-truth dataset. 120 Craigslist advertisements, with 20 randomly selected from each study region, were manually annotated by $3$ human annotators. Each annotator reads and annotates the 120 advertisements independently. Thus, each advertisement receives annotations from $3$ human judges. We then adopt a rule of majority vote, and the place names which are identified by at least two annotators are kept, while those labeled by only one annotator are discarded. The obtained 120 annotated advertisements are used as the ground truth for evaluation experiments\footnote{The dataset is available at: \url{https://github.com/YingjieHu/LocalPlaceName}}. While this dataset is a small sample, it nevertheless enables us to quantitatively measure the performances of the proposed framework. Further evaluation experiments can be conducted when more human-annotated advertisements have become available.

To quantify the performance of a model, we employ three metrics from information retrieval, which are precision, recall, and F-score (Equations 4 to 6).
\begin{eqnarray}
Precision = \frac{|Retrieved \: Relevant|}{|All \: Retrieved|} \\
Recall = \frac{|Retrieved \: Relevant|}{|All \: Relevant|} \\
\fscore = 2\cdot\: \frac{Precision \times Recall}{Precision + Recall}
\end{eqnarray}
Precision measures the percentage of correctly identified place names among all the names returned by a model. Recall measures the percentage of correctly identified place names among all the place names that should be identified (i.e., the ground-truth place names labeled out by human judges). F-score is the harmonic mean of precision and recall. F-score is high when both precision and recall are fairly high, and is low if either of the two is low.

With the ground-truth data and the evaluation metrics, we first quantify the performances of the four NER models in Stage 1. Evaluating the performance of each stage can help us understand the functioning details of the entire framework. We test the performance of each individual NER model (Table \ref{individual_NER}), and the performances of the combined models (Table \ref{ner_combined}).
\begin{table}[h]
	\centering
	\caption{Performance of each NER model.}
	\vspace*{0.3cm}
	\label{individual_NER}
	\begin{tabular}{|c|c|c|c|c|}
		\hline
			& spaCy          & Stanford       & \begin{tabular}[c]{@{}c@{}}Stanford \\  Case-insensitive\end{tabular} & \begin{tabular}[c]{@{}c@{}}Stanford \\  Twitter-retrained\end{tabular} \\ \hline
		Precision & 0.396          & 0.570          & 0.536             & 0.451            \\ \hline
		Recall    & \textbf{0.663} & \textbf{0.672} & \textbf{0.522}    & \textbf{0.668}   \\ \hline
		F-score   & 0.496          & 0.617          & 0.529             & 0.538            \\ \hline
	\end{tabular}
\end{table}

\begin{table}[h]
	\centering
	\caption{Performances of the combined NER models.}
	\vspace*{0.3cm}
	\label{ner_combined}
	\begin{tabular}{|c|c|c|c|c|}
		\hline
		& spaCy          & \begin{tabular}[c]{@{}c@{}}spaCy \\ + Stanford\end{tabular} & \begin{tabular}[c]{@{}c@{}}Former 2 \\ + Case-insensitive\end{tabular} & \begin{tabular}[c]{@{}c@{}}Former 3 \\ + Twitter-retrained\end{tabular} \\ \hline
		Precision & 0.396          & 0.399                                                       & 0.377                                                                   & 0.336                                                                  \\ \hline
		Recall    & \textbf{0.663} & \textbf{0.839}                                              & \textbf{0.864}                                                          & \textbf{0.932}                                                         \\ \hline
		F-score   & 0.496          & 0.541                                                       & 0.525                                                                   & 0.494                                                                  \\ \hline
	\end{tabular}
\end{table}

The goal of Stage 1 is to identify as many place name candidates as possible. Thus, we focus on the evaluation metric of recall. As can be seen from these two tables, using the NER models individually achieves a recall from $0.522$ to $0.672$, while combining the four models gives us a much higher value, $0.932$. It is worth noting that similar recall values do not mean that the NER models extract the almost same set of place name candidates. For example, spaCy NER and the default Stanford NER have a recall of $0.663$ and $0.672$ respectively. A combination of the two produces a recall of $0.839$, suggesting that each NER model extracts certain place name candidates which are not identified by the other model. By combining multiple NER models, we can identify more place name candidates (thus, higher recall). Meanwhile, such a union combination introduces more noise terms into the output of Stage 1 (thus, lower precision).

\begin{figure}[h]
	\vspace*{-0.5cm}
	\begin{center}
		\includegraphics[width=0.9\textwidth]{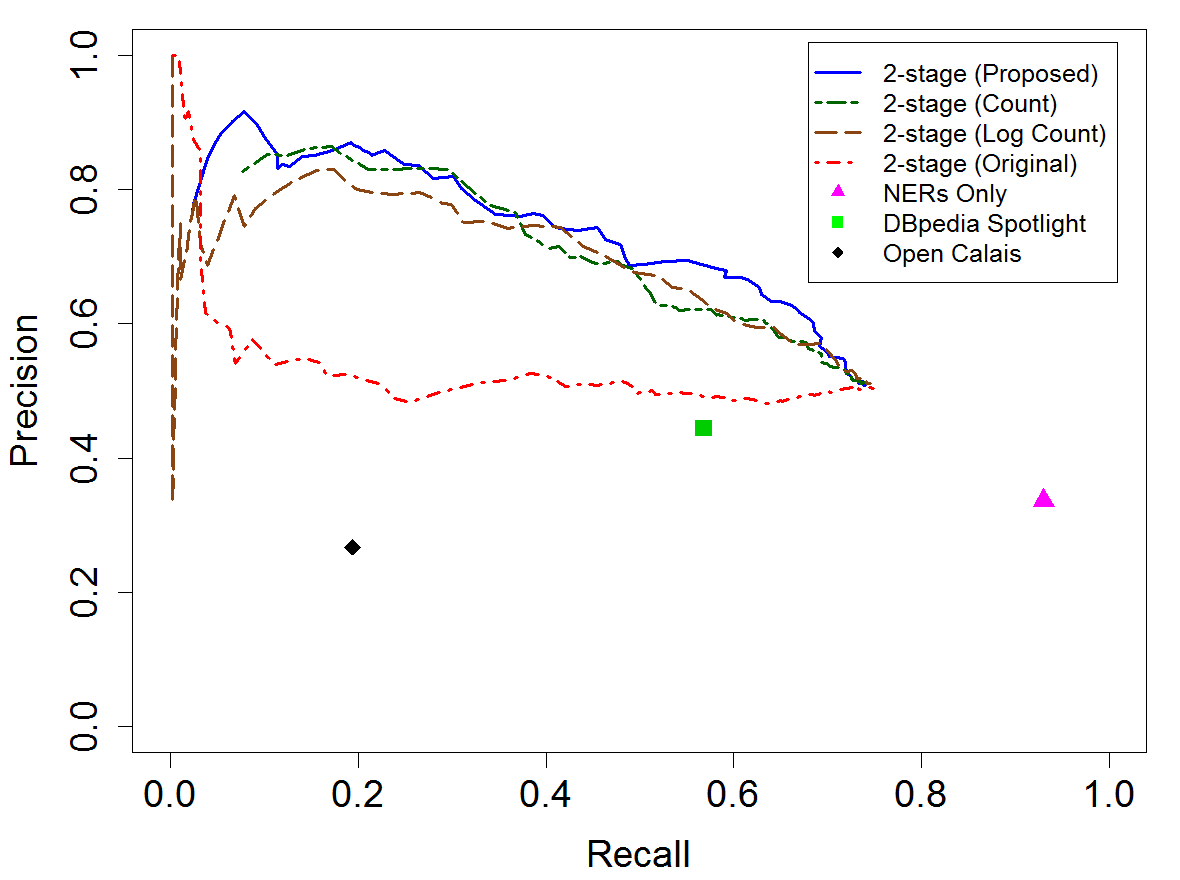}
		\caption{Performances of the proposed  framework and the $6$ baseline models.}
		\label{performance}
	\end{center}
	\vspace*{-0.5cm}
\end{figure}

We continue to evaluate the performance of Stage 2, which also determines the final performance of the framework. The goal of Stage 2 is to ``weed out" the false positives included in the output of Stage 1. With different thresholds for the adjusted entropy sum, different sets of place name candidates can be returned as the final output of Stage 2. We normalize the adjusted entropy sums into $[0,1]$ based on the minimum and maximum values, and iterate the threshold from $0$ to $1$ with a step $0.01$. At each threshold, we can obtain a precision, a recall, and a F-score. Using recall as the $x$ coordinate and precision as the $y$ coordinate, we can plot out a precision-recall curve to show the performances of Stage 2 (also the entire framework) at different thresholds. Figure \ref{performance} shows the precision-recall curve  of our proposed two-stage framework (the blue curve). 
For comparison, we also plot out the precision-recall curves of the original SSI (without adjusting the entropy sums using the point counts) and two other possible ways for adjusting the entropy sums (i.e., point count, $1/|L_x|$, and log point count, $1/\log(|L_x|)$). In addition, we compare our framework with two other NER models, DBpedia Spotlight \cite[]{daiber2013improving} and Open Calais (\url{http://www.opencalais.com}), which extract only the entities contained in Wikipedia. The performance of Stage 1 (NERs only) is also plotted for comparison.


It is easy to notice that the performances of some models are represented as points while those of others are as curves in Fig. \ref{performance}.  DBpedia Spotlight and Open Calais are represented as points, because they are off-the-shelf tools which directly output the recognized entities. Thus, their results are quantified by one precision and one recall. A similar situation applies to the combined NER models of Stage 1 (the pink triangle in the figure). The results of the two-stage models are represented as curves, because different thresholds can be used to control the returned place name candidates. Thus, there is one precision and one recall at each threshold, which allow the plotting of curves. 

We can evaluate the proposed two-stage framework by comparing its precision-recall curve with the performances of the other models. Overall, our framework outperforms DBpedia Spotlight and Open Calais in both precision and recall. This result suggests that our framework can correctly identify more local place names than the two NER models that rely on an existing knowledge base. In addition, the 3 modified SSI models all outperform the original SSI which does not adjust the entropy sums. In particular, the proposed adjustment (using $1/\sqrt{|L_x|}$) shows a better performance than the two tested alternatives (using $1/|L_x|$ or $1/\log(|L_x|)$).

We determine a threshold for the adjusted entropy sum to generate the final output. Such a threshold can be determined based on  application needs. An application that favors a high percentage of correctness in the output can use a threshold that produces high precision (but low recall). In contrast, an application that favors high coverage of the result can use a threshold that gives high recall (but low precision). For a balanced performance, one can choose the threshold with the highest F-score. In this work, both precision and recall are important, but we favor recall slightly more since a higher recall allows us to extract more local place names. Thus, we determine the threshold using the following procedure: 1) we rank the thresholds based on their F-scores, and identify those that achieve the top $10$ F-scores; 2) among the $10$ thresholds, we identify the one that produces the highest recall. Using this procedure, we select a final threshold $0.67$ which gives a precision $0.600$, a recall $0.684$, and a F-score $0.639$. Generally, false positives happen when a non-place term also shows certain clusterness based on the geotagged advertisement locations. For example, the names of some realtors are included in the final result. By examining the associated locations, we find that these points do cluster at some neighborhoods, since a realtor is often in charge of one or several neighborhoods. On the other hand, false negatives often happen when a place name is not mentioned by enough housing advertisements. These place names are directly removed during the filtering process since they are associated with only one or two points.

We also compare the final performance of the two-stage framework with using Stage 1 alone. Using a combination of the four NER models, Stage 1 achieves a precision $0.336$ and a very high recall $0.932$. Adding Stage 2 increases the precision to $0.600$ but also decreases the recall to $0.684$. In an ideal case, Stage 2 will only weed out the false positives, and thus we should see an increased precision and no-decrease (or slightly-decreased) recall. In practice, however, Stage 2 also removes some true place names in the weeding-out process. While Stage 2 has decreased the recall, it largely reduces the number of place name candidates included in the output. To give a concrete example, the output of Stage 1 for the Los Angeles study region contains $27823$ place name candidates, while only $832$ names are kept after Stage 2. Thus, Stage 2 largely filters out the false positives, although it also mistakenly removes some true place names. For the purpose of enriching gazetteers, the output of Stage 1 contains too many noise terms and cannot be directly utilized. While the output of the two-stage framework also contains some non-place names, it is feasible to go through the relatively small numbers of extracted place names, and identify the false positives that have slipped away from Stage 2. Table \ref{table_totalNumber} shows the total numbers of terms (with both true place and non-place names) identified by the two-stage framework for the study regions. 

\begin{table}[h]
   	\centering
   	\caption{Counts of terms extracted for the 6 study regions.}
   	\vspace*{0.3cm}
   	\label{table_totalNumber}
   	\begin{tabular}{|c|c|c|c|c|c|}
   		\hline
   		New York & Los Angeles & Chicago & Richmond & Boise & Spokane \\ \hline
   		408      & 832         & 448     & 239      & 222   & 178     \\ \hline
   	\end{tabular}
\end{table}

\subsection{Comparison with existing gazetteers}
One important goal of the proposed framework is to enrich existing gazetteers with additional place names, especially local place names. In this subsection, we compare the place names extracted for the six regions with the place names in four existing gazetteers, which are Foursquare venues, GeoNames, TGN, and WOF. Foursquare venues are maintained by the location-based social media company Foursquare, but a vast majority of place entries are contributed by its users. GeoNames is a combination of multiple existing gazetteers from both authorities and commercial companies.
TGN  represents the gazetteers developed by a single authority which uses strict editorial rules to control the included place entries. WOF is an open gazetteer from Mapzen with place entries selected from a variety of sources such as Quattroshapes and Natural Earth.

Additional considerations are necessary for comparing the extracted place names with the existing gazetteer entries. As discussed previously, the output place names from our experiments still contain non-place names. Thus, simply counting the number of extracted place names that do not have counterparts in the gazetteers can result in an overestimation. To provide a more robust estimation, we count only those names that are indeed place names. To do so, we perform the following steps using Foursquare venues.
\begin{enumerate}
	\item Compare each extracted place name with Foursquare, and identify the place names that have a direct match (case insensitive) in Foursquare. For example, ``ann morrison park" from our output is a direct match with ``Ann Morrison Park" in Foursquare.
	\item Compare the rest of the extracted place names with Foursquare, and identify the place names that are indirect matches with Foursquare entries. For example, ``bsu" from our output is an indirect match with ``Boise State University (BSU) Education Building" in Foursquare.
	\item For the rest of the extracted place names, we verify each by searching online to determine whether it is indeed a place name.
\end{enumerate}

We use Foursquare venues (instead of the other gazetteers), because it is designed to provide a search-and-discovery service for local places. Meanwhile, local users also contribute many place entries to Foursquare. Thus, we expect that Foursquare contains the most comprehensive local places among the compared gazetteers. As a result, we can minimize the number of place names that need to be manually verified. Strict rules are also used in step (3) to eliminate certain place names, which are as follows:
\begin{itemize}
	\item Small streets in an apartment complex are not considered as new place entries. For example, we identified $11$ small streets for Spokane, which are not contained in Foursquare venues. Such small streets are not counted as the discovered new place names.
	\item Alternative spellings, which simply add or remove spaces, are not counted as new place entries. For example, ``Green Belt" is used in multiple housing advertisements in Boise, but it is not considered as a new place name, since the proper spelling ``Greenbelt" is already included in Foursquare. However, alternative names, such as ``K-Town" for ``Koreatown", is counted as a discovered new place name.  
\end{itemize}
The above rules are adopted to help generate a robust estimation on the number of new place names discovered by our framework. These rules, however, are not meant to be fixed and can be adjusted based on practical needs (e.g., one could also consider ``Green Belt" as a valid local name in an application).

The numbers of discovered new place names in comparison with Foursquare are reported in Table \ref{table_gazetteer_comparison}. To compare with GeoNames, TGN, and WOF, we count the extracted place names that do not have any match (both direct and indirect matches) in these three gazetteers but have a direct match in Foursquare or are verified as true place names in step (3).  
\begin{table}[h]
	\centering
	\caption{Estimated numbers of new local place names discovered using the proposed framework in comparison with four existing gazetteers.}
	\vspace*{0.2cm}
	\label{table_gazetteer_comparison}
	\begin{tabular}{|c|c|c|c|c|}
		\hline
		& Foursquare & GeoNames & TGN & WOF \\ \hline
		New York    & 3          & 51       & 148 & 99 \\ \hline
		Los Angeles & 6          & 159      & 330 & 175 \\ \hline
		Chicago     & 3          & 75       & 134 & 81\\ \hline
		Richmond    & 6          & 53       & 81  & 56 \\ \hline
		Boise       & 2          & 59       & 68  & 58 \\ \hline
		Spokane     & 2          & 20       & 45  & 38\\ \hline
	\end{tabular}
\end{table} 

\begin{figure}[h]
	\vspace*{-0.5cm}
	\begin{center}
		\includegraphics[width=\textwidth]{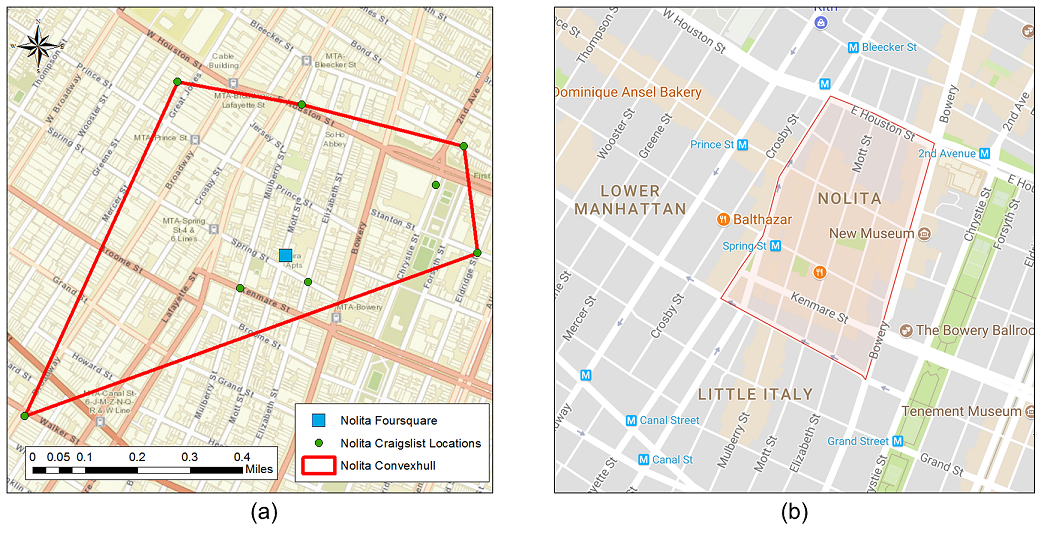}
		\caption{(a) Convex hull of ``Nolita" constructed based on housing advertisement locations and the point representation of ``Nolita" in Foursquare; (b) ``Nolita" on Google Maps.}
		\label{nolita_google}
	\end{center}
	\vspace*{-0.5cm}
\end{figure}

Compared with Foursquare venues, the proposed framework only discovered a handful of new local place names. As a location-based social media and a local search service provider, Foursquare already contains many place names extracted by our approach. The small number of new place names not contained in Foursquare include districts, such as ``Bell School District" in Chicago and ``Central Business District" in Richmond. We also discovered quite a few alternative place names, including ``K-Town" for ``Koreatown", ``Lamplighter Coffee" for ``Lamplighter Roasting Co.", and ``Bio Park" for ``Virginia BioTechnology Research". While our approach does not extract many new places compared with Foursquare, it has values in three aspects. First, Foursquare dataset is a commercial product which has usage restrictions, while our place names are extracted from publicly available local housing advertisements using open methods.  Second, some geographic regions may have very few or no Foursquare users, while housing advertisements can be found in almost anywhere people live. Third, Foursquare provides only point representations for most place names, while our approach allows the construction of rough spatial footprints. Figure \ref{nolita_google}(a) shows the convex hull constructed based on the housing advertisement locations associated with ``Nolita" (for ``North of Little Italy") in New York City, while Foursquare represents ``Nolita" as one point.
Figure \ref{nolita_google}(b) shows the boundary of ``Nolita" on Google Maps. It is interesting to see that the convex hull does not exactly match the boundary of ``Nolita" on Google Maps, but includes some housing properties that seem to be outside of ``Nolita". One possible reason is that advertisement writers may describe their properties as ``nearby" or even ``within" a neighborhood so that the advertised housing property could become more attractive to potential buyers or renter. Such a phenomenon was also found by researchers of \textit{The Neighborhood Project} (\url{https://hood.theory.org}). On the other hand, there is no guarantee that the boundaries on Google Maps are absolutely correct, since neighborhood boundaries are usually fuzzy \mbox{\cite[]{greene2012exploring}}. 

In comparison with GeoNames, TGN, and WOF, our method discovers considerable numbers of new place names ranging from $20$ to $330$. It can be seen from Table \mbox{\ref{table_gazetteer_comparison}}, more place names are discovered for TGN than for GeoNames or WOF. Such a result is understandable since TGN was designed to store place names often with important historical meaning rather than local or informal place names. In addition, more names are identified for the first three study regions containing megacities than for the other three regions which contain smaller cities. Below we list some example place names that are discovered by our approach:
\begin{itemize}
	\item \textbf{Local neighborhoods}: ``Hyde Park" in Boise, ``West Loop" in Chicago, ``Silicon Beach" in Los Angeles, ``Museum District" in Richmond.
	\item \textbf{Parks}: ``Elm Grove Park" in Boise, ``Pan Pacific Park" in Los Angeles, ``Deep Run Park" in Richmond.
	\item \textbf{Schools}:  ``Loyola Law School" in Los Angeles, ``Sawtooth Middle School" in Boise, ``Prairie View Elementary" in Spokane.
	\item \textbf{Points of interest}: ``Plum Market" in Chicago, ``Barclay Center" in New York, ``Howard Hughes Center" in Los Angeles.
	\item \textbf{Alternative names}: ``FiDi" in New York, ``Central Bench" in Boise, ``DTLA" in Los Angeles.
\end{itemize}

\section{Conclusions and future work}

Local place names can support important geospatial applications in disaster response, urban planning, and many other areas. This paper presents a two-stage computational framework for extracting local place names in a given geographic region based on geotagged housing advertisements posted on local-oriented websites, such as Craigslist. The first stage of the framework focuses on the textual content of the advertisements, and uses a combination of off-the-shelf and retrained named entity recognition models to identify place name candidates from the text. The second stage examines the point locations associated with each place name candidate, and uses a geospatial clustering algorithm, modified  scale-structure identification, to quantify the geo-indicativeness of the place name candidates. A threshold can then be decided to filter out non-place names. We applied the proposed two-stage framework to geotagged housing advertisements in six different regions, and evaluated its performances in terms of precision, recall, and F-score. We also compared the extracted place names with the entries in four existing gazetteers, which are Foursquare venues, GeoNames, TGN, and WOF, to demonstrate the local place names discovered by the proposed framework. 

The contributions of this paper can be seen from two perspectives. From the perspective of application, this paper presents an innovative use of geotagged housing advertisements for extracting local place names. This type of data contains local place names, is widely available, and captures newly-constructed geographic entities. From the perspective of methodology, this work presents an integration of natural language processing and geospatial clustering methods. As indicated by the experiment results, integrating geospatial clustering with NLP methods has a better performance in extracting local place names than using the methods based on linguistic features alone. 

The proposed framework has its limitations and can be improved in the near future. First, the final output of the framework still contains quite some non-place names. Some terms slipped through the filtering process of Stage 2, because they show certain geo-indicativeness similar to those of the true place names (e.g., realtor names). Further studies can be conducted on removing these and other false positives. Second, deeper natural language analysis can be performed on the textual descriptions of the housing advertisements. For example, we can differentiate the housing advertisements which state ``... is located \textit{within} Nolita" from those which state ``... is located \textit{close to} Nolita" in order to obtain a more accurate spatial footprint of the place name. While further research can be conducted, we hope that this paper has made a modest contribution to harvesting local place names.


\section*{Acknowledgments}
The authors would like to thank the three anonymous reviewers for their constructive suggestions and comments. This research is supported by the Professional and Scholarly Development Award (Award Number: R011038-002) from the University of Tennessee, Knoxville.

\bibliographystyle{tGIS}
\bibliography{reference}

\vspace{36pt}


\label{lastpage}

\end{document}